\documentclass[10pt]{article}
\usepackage{epsfig,graphicx}
\usepackage{times,ch2005}

\def\gtrsim{\lower.5ex\hbox{$\; \buildrel > \over\sim \;$}} 
\def\lesssim{\lower.5ex\hbox{$\; \buildrel < \over\sim \;$}}

\def\beq{\begin{equation}}
\def\eeq#1{\label{#1}\end{equation}}
\def\eeqn{\end{equation}}

\def\beqa{\begin{eqnarray}}
\def\eeqa#1{\label{#1}\end{eqnarray}}
\def\eeqan{\end{eqnarray}}

\let\bar=\overbar

\def\Dslash{\not{\hbox{\kern-4pt $D$}}}
\def\dslash{\not{\hbox{\kern-2pt $\del$}}}

\newcommand{\tev}{\ensuremath{\mathrm{\,Te\kern -0.1em V}}\xspace}
\newcommand{\gev}{\ensuremath{\mathrm{\,Ge\kern -0.1em V}}\xspace}
\newcommand{\mev}{\ensuremath{\mathrm{\,Me\kern -0.1em V}}\xspace}
\newcommand{\kev}{\ensuremath{\mathrm{\,ke\kern -0.1em V}}\xspace}
\newcommand{\ev}{\ensuremath{\mathrm{\,e\kern -0.1em V}}\xspace}
\newcommand{\gevc}{\ensuremath{{\mathrm{\,Ge\kern -0.1em V\!/}c}}\xspace}
\newcommand{\mevc}{\ensuremath{{\mathrm{\,Me\kern -0.1em V\!/}c}}\xspace}
\newcommand{\gevcc}{\ensuremath{{\mathrm{\,Ge\kern -0.1em V\!/}c^2}}\xspace}
\newcommand{\mevcc}{\ensuremath{{\mathrm{\,Me\kern -0.1em V\!/}c^2}}\xspace}

\def\mus  {\ensuremath{\rm \,\mus}\xspace}

\def\mus        {\ensuremath{\,\mu{\rm s}}\xspace}    

\begin{document}


\Title{Gamma Ray Bursts as Possible High Energy Sources}
\bigskip


%
\label{Dermer}

%
\author{Charles Dermer\index{Dermer, C.} }

%
\address{U.S. Naval Research Laboratory\\
Code 7653, 4555 Overlook Ave., SW \\
Washington, D.C. 20375-5352 USA \\
}

\makeauthor\abstracts{
Gamma-ray bursts are known to be sources of high-energy $\gamma$ rays,
and are likely to be sources of high-energy cosmic rays and neutrinos.
Following a short review of observations of GRBs at multi-MeV energies
and above, the physics of leptonic and hadronic models of GRBs is
summarized. Evidence for two components in BATSE and EGRET/TASC data
suggest that GRBs are sources of high-energy cosmic rays. GLAST
observations will reveal the high-energy $\gamma$-ray power and energy
releases from GRBs, and will provide detailed knowledge of anomalous
high-energy emission components, but confirmation of cosmic ray
acceleration must await 100 TeV -- PeV neutrino detection from GRBs.
}

\section{Introduction}
Gamma-ray burst (GRB) studies represent one of the most dynamic fields
in contemporary astronomy, and the field continues to evolve rapidly
as new instrumentation comes online. Knowledge of GRBs in the
high-energy, multi-MeV regime is poised to undergo great advances in
the near future.  Right now, Swift is exploring the late prompt and
early afterglow phases of GRBs at X-ray and hard X-ray energies with
unprecedented detail. The ground-based air Cherenkov telescopes,
including HESS, VERITAS, and MAGIC, are reaching better sensitivities
and lower thresholds with the goal of detecting GRBs at $\sim 100$ GeV
-- TeV energies. The MAGIC telescope has already demonstrated the
ability to slew within $\approx 30$ s to a GRB. AGILE, a small
scientific mission developed by the Italian Space Agency, is due for
launch next year. With sensitivity comparable to EGRET, though with a
much larger field-of-view (factor-of-two decline in sensitivity at an
off-axis angle of $\sim 60^\circ$, versus $\sim 25^\circ$ for EGRET),
it should detect some 5 -- 10 GRBs per year above 100 MeV. The GLAST
mission, which includes both the Gamma-ray Burst Monitor and Large
Area Telescope that will cover energies from $\approx 5$ keV and
above, is scheduled for launch in September 2007.  Its peak effective
area at $\approx 1$ GeV will be $\approx 8\times$ greater than EGRET's
at $\approx 100$ MeV, and its field-of-view represents $\approx
1/6^{\rm th}$ of the full sky, compared to $\approx 1/20^{\rm th}$ for
EGRET. This should lead to the detection of some $30$ -- 100 GRBs per
year at $\gtrsim 100$ MeV energies.

Beyond the electromagnetic realm, there is the real possibility that
GRBs will be the first detected sources of high-energy ($\gg 1$ TeV)
neutrinos. The south pole AMANDA neutrino detector is, as new strings
are added, evolving into the km$^3$ IceCube neutrino telescope over
the course of this decade. AMANDA has already detected thousands of
cosmic-ray induced background neutrinos events, and improved
sensitivity and background rejection should soon reveal meaningful
upper limits on GRB source models if not direct detections.  If GRBs
are sources of ultra-high energy cosmic rays (UHECRs), then detection
of $\gtrsim 10^{17}$ eV neutrino coincident with GRBs may be possible
with Auger and ANITA. In view of the rapid technological advances
expected over the next 5 years, therefore, we can expect that our
knowledge of GRBs as high-energy sources will vastly increase.

In this contribution, I summarize what is now known from observations
of high-energy radiation from GRBs. This includes primarily the {\it
Compton Gamma Ray Observatory} results, which have provided the most
significant detections of multi-MeV emission from GRBs. An important
result from EGRET is that there is good evidence for two components in
GRBs, and incontrovertible evidence for an extended phase, as observed
from GRB 940217. In addition, an anomalous $\gamma$-ray emission
component was detected from GRB 941017. There is, moreover, a
tantalizing suggestion of TeV emission from GRB 970417a, made with the
Milagrito telescope.

Interpretation of these results is made within the blast-wave
scenario, which was originally developed to understand GRB afterglows
\cite{mr97}. Although normally used to model leptonic afterglow
emissions from GRBs, this scenario can also be used to predict
emission signatures from hadrons accelerated by GRBs. If GRBs
accelerate UHECRs, then anomalous $\gamma$-ray emission signatures are
expected from GRBs. Studies of high-energy radiation from GRBs have
the potential to answer one of the outstanding questions in
contemporary astronomy, namely the origin of the cosmic rays.
Detection of $\gtrsim 10^{14}$ eV  neutrinos from GRBs would provide
compelling evidence in support of this solution.

\section{Observations of High Energy Radiation from GRBs}

The {\rm Solar Maximum Mission} satellite revealed that $\gtrsim 1$
MeV emission was a common property of GRBs \cite{mea85}, thus
establishing that the radiation has a nonthermal origin.  COMPTEL
detected over 30 GRBs at photon energies $E > 0.75$ MeV
\cite{connorsea98}.  The spark chamber on EGRET detected $\gtrsim 30$
MeV photons from 7 GRBs \cite{dcs98}. These GRBs are invariably among
the most fluent BATSE bursts, indicating that detection of $\gtrsim
100$ MeV emission is sensitivity-limited rather than a property of
some subset of long-duration GRBs.  The average photon spectral index
of four EGRET GRBs (which includes GRB 940217), consisting of 45
photons with $E > 30$ MeV, is $\langle \alpha \rangle = 1.95\pm 0.25$,
consistent with being a high-energy extension of the spectrum observed
with BATSE.

\begin{figure}[t]
\vskip-2.0in
\hskip0.75in\includegraphics[scale=0.45]{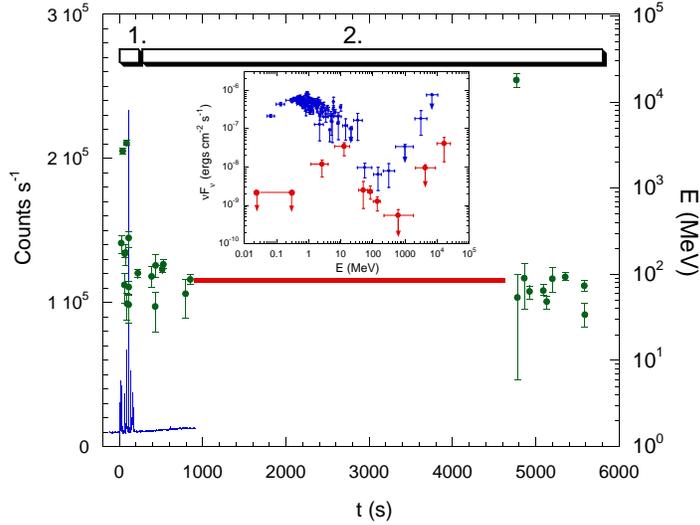}   
\vskip-0.2in  
\caption{BATSE light curve, and times and energies of 
EGRET-detected photons \cite{hea94} of GRB 940217. Inset shows
composite $\nu F_\nu$ spectra during the initial 180 s interval 1
(small blue symbols), and the later interval 2 (large red symbols). }
\end{figure}

Fig.\ 1 shows the light curve and spectra of the famous burst GRB
940217, which displayed an Earth-occulted $\sim 100$ MeV tail that
lasted for $\sim 95$ minutes, two $\sim 3$ GeV photons during the 180
s time interval when the BATSE emission was detected, and an 18 GeV
photon 90 minutes later \cite{hea94}. The total number of spark
chamber events was 28, with 10 photons observed during the first 180
s, and another 10 after Earth occultation. The total $> 20$ keV
fluence of this event over the BATSE energy range was $\gtrsim
6.6\times 10^{-4}$ ergs cm$^{-2}$.

The interval 1 and 2 $\nu F_\nu$ spectra are shown in the inset.  The
interval 1 spectrum clearly shows a second component rising in the 100
MeV -- GeV range with no evidence of a cutoff.  Even during interval
2, which lasted for 5,400 s following the 180 s GRB, there is evidence
from the EGRET Total Absorption Shower Counter (TASC) and spark
chamber for two distinct components.  The TASC, which measured spectra
in the $\sim 1$ -- 200 MeV range and served as a calorimeter to
measure total photon energies for EGRET, detected at least 26 GRBs
\cite{cds98,gon03}.  Joint analysis \cite{gon03} of the BATSE Large
Area Detector (LAD) and the EGRET TASC data resulted in the detection
of an anomalous MeV emission component in the spectrum of GRB 941017
that decays more slowly than the prompt emission detected with the LAD
in the $\approx 50$ keV -- 1 MeV range. The multi-MeV component lasted
for $\gtrsim 200$ seconds (the $t_{90}$ duration of the lower-energy
prompt component was 77 sec), and was detected with the LAD near 1 MeV
and with the TASC between $\approx 1$ and 200 MeV. The spectrum is
very hard, with a photon number flux $\phi(E)\propto E^{-1}$.
Anomalous emission components have now been detected in at least two
other GRBs (M.\ M.\ Gonz\'alez and B.\ L.\ Dingus, private
communication, 2004).

At TeV energies, analysis \cite{atk00} of data from the all-sky water
Cherenkov telescope Milagrito correlated with the times of 54 BATSE
GRBs within its field of view resulted in one statistically
significant excess, namely GRB 970417a.  This BATSE GRB had a
relatively low fluence of $\approx 1.5\times 10^{-7}$ ergs cm$^{-2}$
at 50 -- 300 keV energies, and a $\nu F_\nu$ peak energy $E_{pk}$
below the BATSE energy band, so that it would technically be described
as an X-ray flash or X-ray rich GRB. Besides needing to be relatively
nearby (redshift $z \lesssim 0.2$) to avoid strong attenuation on the
diffuse intergalactic infrared radiation field, its TeV fluence would
have to be at least an order of magnitude greater than the BATSE
fluence, i.\ e., $\gtrsim 10^{-6}$ ergs cm$^{-2}$, to be detected with
Milagrito.

No evidence for high-energy neutrinos coincident with GRBs has yet
been reported with the AMANDA array \cite{ice05}.

\section{Models for High Energy $\gamma$ Rays from GRBs}

We consider both leptonic and hadronic models for emissions from GRBs,
focusing on the particular cases of GRB 940217, GRB 970417a, and GRB
941017.

\subsection{GRB 940217}

The discovery of the extended phase of emission from GRB 940217 took
place prior to the confirmation that GRBs originate from cosmological
distances.  Even now, however, there is no agreed explanation for this
high-energy radiation. An early model by Katz \cite{kat94} argued that
collisions of cosmic rays with gas in a dense clouds produced
secondary $\pi^0$ radiation. This model cannot possibly be valid for a
source at cosmological distances, given the required energies and
decay time scale. A second early model, by M\'esz\'aros and Rees
\cite{mr94}, is sketched within the context of a relativistic outflow
from a cosmological source. Here, the delayed emission is due either
to successive waves of ejecta that collide at later times, or in terms
of relativistic ejecta being slowed by the external medium to form GeV
radiation.  With the suggestion that UHECRs are accelerated by GRB
blast waves \cite{vie95,wax95}, it was recognized that cascade
radiation signatures of UHECRs interacting with photons of the
microwave background could produce GeV--TeV photons \cite{wc96}. This
could not, however, account for the emission from GRB 940217, which is
correlated with peaks in the BATSE emission during the prompt phase,
and decays on a much shorter timescale than would occur from cascade
processes involving diffuse intergalactic radiation.

\begin{figure}[t]
\vskip-3.3in
\hskip0.2in\includegraphics[scale=0.50]{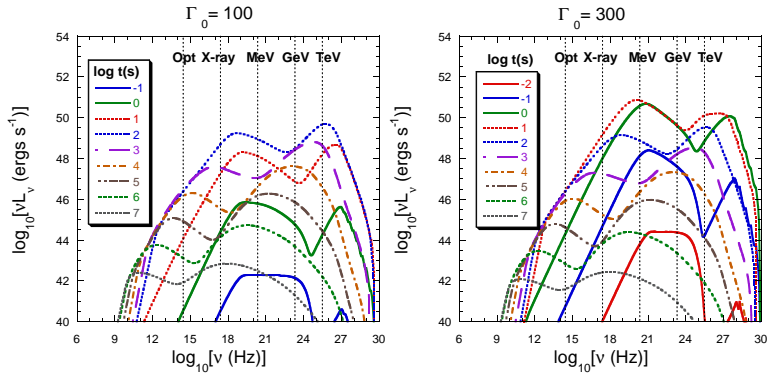}    
\caption{Calculations of SEDs from uncollimated GRB blast waves 
that are energized, decelerate and radiate by capturing material from
a uniform surrounding medium; parameters are described in the text.
The initial Lorentz factor $\Gamma_0 = 100$ and $\Gamma_0 = 300$ in
the left  and right  panels, respectively. The $\nu L_\nu$ SEDs are
shown at observer times of $10^j$ seconds after the onset of the GRB
event, with $j$ given in the captions.}
\end{figure}

Progress in the development of the blast-wave model of GRBs permitted
a quantitative application of this model to observations of
high-energy radiation from GRBs.  Using a numerical simulation code
developed by J. Chiang \cite{cd99}, the synchrotron and synchrotron
self-Compton (SSC) emissions in a leptonic blast wave model were
modeled, giving the results shown in Fig.\ 2.  In addition to
synchrotron and SSC processes, the numerical simulation model
\cite{dcm99} includes synchrotron self-absorption and adiabatic loss
processes, and follows blast-wave evolution self-consistently.  The
photons are attenuated by internal $\gamma\gamma$ absorption, but pair
reinjection is not followed.

In this calculation, we consider an external shock model for a GRB
taking place in a uniform surrounding medium with density $n_0 = 100$
cm$^{-3}$. The GRB is assumed to produce a thin blast wave where only
the forward shock produces strong particle acceleration and radiation.
The apparent isotropic explosion energy of the GRB is $10^{54}$ ergs,
and the fraction $\epsilon_e $ of nonthermal swept-up proton kinetic
energy transferred to nonthermal electrons is 0.5, the injection index
$p = 2.5$ for the electrons, and $\epsilon_B = 10^{-4}$, where the
comoving magnetic field strength $B$ is defined by the relation
$B^2/8\pi = 4\epsilon_B \Gamma^2 m_pc^2 n_0$.  The maximum electron
Lorentz factor is determined by the the value where the gyration
timescale equals the synchrotron energy-loss timescale.  Fig.\ 2 shows
temporally evolving $\nu L_\nu$ spectra for an uncollimated blast wave
with two values of $\Gamma_0$.  The microscopic parameters are assumed
to be time-independent.

In both calculations, the SSC process makes a distinct component that
rises at multi-GeV energies to produce emission at TeV energies before
being attenuated by internal absorption.  The $\gamma\gamma$ process
degrades only $\gtrsim$ 1 TeV photons in the models shown here. Thus
the internal attenuation of high-energy gamma rays is not too severe
and the SSC component is bright enough that TeV radiation is produced
at a comparable $\nu F_\nu$ level as the synchrotron radiation. By
studying the temporal dependence of the $\sim $ GeV radiation in the
$\Gamma_0 = 300$ case, one sees that because of the deceleration of
the blast wave, the SSC flux sweeping through the EGRET band could in
principle make an emission component over a few thousand seconds that
would decay much more slowly than the X-ray and MeV emissions.  We
proposed \cite{dcm99} that this was the origin of the extended
emission component in GRB 941017. Careful examination of the numerical
simulation result for the $\Gamma_0 = 300$ case shows, however, that
there is at least a one-to-two order-of-magnitude decline in the GeV
flux over the first several thousand seconds. Suitable choice of
parameters, for example, by reducing the maximum electron energy,
could make a relatively constant $\approx 100$ MeV -- GeV flux for an
hour or two, so this model provides a viable explanation for the
delayed emission from GRB 941017.

\subsection{GRB 970417a}

The lower Lorentz factor (``dirty fireball") case with $\Gamma_0 =
100$ in Fig.\ 2 produces a larger relative $\nu F_\nu$ flux in the SSC
component than in the synchrotron component and, moreover, produces
the $E_{pk}$ value of the synchrotron component at keV rather than MeV
energies.  The TeV radiation detected by Milagro from GRB 970417a
could originate from the SSC emission from a nearby $z\lesssim 0.2$
GRB \cite{dcm99}.

Although the leptonic blast wave model is consistent with observations
of the TeV emission from GRB 970417a, other explanations are possible,
including proton synchrotron radiation \cite{fra04}. This alternative
would require very strong magnetic fields in the GRB blast waves, with
energy requirements that may be unacceptable.  Another important issue
that will arise if TeV emission is confirmed from GRBs is whether such
emission is compatible with an internal shock model. The
$\gamma\gamma$ absorption at TeV emission is very large unless the
bulk Lorentz factors $\Gamma_0$ of the outflows exceed several
hundreds \cite{raz04}.
 
\subsection{GRB 941017}

In contrast, an external shock model with strong forward shock
emission does not easily explain the anomalous $\gamma$-ray emission
component from GRB 941017.  It has been proposed that within the
standard leptonic blast-wave model, this separate component could be
reverse-shock emission Compton-scattered by the forward shock
electrons \cite{gg03}, including self-absorbed reverse-shock optical
synchrotron radiation \cite{pw04}. This latter model requires very
large apparent isotropic energies exceeding $10^{54}$ ergs, and
forward and reverse shock microphysical parameters that are very
different. Another possibility is that hadronic acceleration in GRB
blast waves could be responsible for this component.

\begin{figure}[t]
\hskip1.in\includegraphics[scale=0.35]{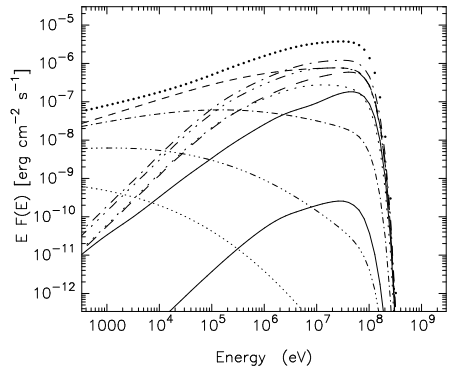}  
\vskip-0.1in   
\caption{Photon energy fluence from an electromagnetic cascade initiated
by photopion secondaries in a lepton/hadronic GRB model, with
parameters given in Ref.\ \cite{da04}. Five generations of Compton
(heavy curves) and synchrotron (light curves) are shown. The first
through fifth generations are given by solid, dashed, dot-dashed,
dot-triple--dashed, and dotted curves, respectively. The total cascade
radiation spectrum is given by the upper bold dotted curve. }
\end{figure}

We have argued \cite{da04} that the anomalous $\gamma$-ray emission
component in GRB 941017 could be a consequence of the acceleration of
hadrons at the relativistic shocks of GRBs. A pair-photon cascade
initiated by photohadronic processes between high-energy hadrons
accelerated in the GRB blast wave and the internal synchrotron
radiation field produces an emission component that appears during the
prompt phase, as shown in Fig.\ 3. Photomeson interactions in the
relativistic blast wave also produce a beam of UHE neutrons, as
proposed for blazar jets \cite{ad03}. Subsequent photopion production
of these neutrons with photons outside the blast wave will produce a
directed hyper-relativistic electron-positron beam in the process of
charged pion decay and the conversion of high-energy photons formed in
$\pi^0$ decay. These energetic leptons produce a synchrotron spectrum
in the radiation reaction-limited regime extending to $\gtrsim$ GeV
energies, with properties in the 1 -- 200 MeV range similar to that
measured from GRB 941017. GRBs displaying these anomalous $\gamma$-ray
components are most likely to be detected as sources of high-energy
neutrinos \cite{gue04}.

A hadronic origin for high-energy $\gamma$-ray emission components is
also attractive in view of the extended phase of high-energy radiation
from GRB 940217. This is because hadronic secondary radiation decays
more slowly than leptonic emission, primarily due to the less
efficient cooling of protons than electrons \cite{bd98}. A proton
synchrotron model faces, however, severe difficulties because of the
large field and particle energies required to accelerate protons to
sufficiently high energies where this process becomes important.  On
the other hand, protons which radiate via photo-meson processes
require dense internal radiation fields that produce strong
attenuation at GeV -- TeV energies. A leptonic origin is more probable
if $\gamma$ ray fluxes extending smoothly to $\gg $ GeV energies is
observed.  If strong anomalous components showing strong
$\gamma\gamma$ attenuation cutoff at energies $\gtrsim 100$ MeV -- few
GeV are observed, as in Fig.\ 3, then this will provide strong evidence
for hadronic acceleration by GRBs.  AGILE and GLAST will therefore
provide crucial evidence to distinguish the leptonic and hadronic
origins of high-energy $\gamma$-ray emissions in GRBs.  An inverse
correlation is expected between neutrino detection and $\gamma$-ray
detection, because a strong internal radiation field, and therefore
also strong $\gamma\gamma$ absorption, is required for efficient
high-energy neutrino production \cite{da03,raz04}.

\section{Cosmic Rays and Neutrinos from GRBs}
 
Because high-energy $\gamma$ rays are emitted by GRBs, high-energy
particles must be accelerated by GRB blast waves. GRBs provide a very
attractive solution to the origin of high-energy cosmic rays (HECRs;
$\gtrsim 10^{14}$ eV) ranging from below the knee of the cosmic-ray
spectrum to the highest energies exceeding $10^{20}$ eV.  For
instance, the luminosity density in $\gamma$ radiation is comparable
to the power required to accelerate super-GZK cosmic rays with
energies exceeding $\cong 6\times 10^{19}$ eV \cite{vie95,wax95}.
GRBs are associated with Type 1c supernovae, and are therefore related
to the most probable accelerators of GeV -- TeV cosmic rays, namely
supernovae of all types.  GRBs are found both in our Galaxy and
throughout the universe; thus they can account for cosmic rays with
energies between $\approx 10^{14}$ eV and $\approx 5\times 10^{17}$ eV
due to GRBs in our Galaxy, and for metagalactic UHECRs due to
extragalactic GRBs \cite{wda04}.

We have recently argued that HECRs originate from galactic and
extragalactic GRBs \cite{wda04}.  In our model, relativistic outflows
in GRBs are assumed to inject power-law distributions of cosmic rays
to the highest ($\gtrsim 10^{20}$ eV) energies.  A diffusive
propagation model for HECRs from a single recent GRB within $\approx
1$~kpc from Earth that took place within the last 0.5 million years
explains the KASCADE data for the cosmic-ray ion spectra near and
above the knee. The cosmic-ray spectrum at energies above the second
knee at $\approx 10^{17.6}$~eV is fit with CRs from extragalactic
GRBs. UHECRs produced by extragalactic GRBs lose energy from momentum
redshifting. Attenuation features in the UHECR flux are produced at
characteristic energies $\sim 4\times 10^{18}$~eV and $\sim 5\times
10^{19}$~eV due to photo-pair and photo-pion energy-loss processes,
respectively.

A fit to the combined KASCADE, HiRes-I and HiRes-II monocular data
between $\approx 8\times 10^{14}$~eV and $3\times 10^{20}$~eV is shown
in Fig.\ 4, with a cosmic-ray number injection index $p=2.2$, an
exponential cutoff energy of $E_{max}=10^{20}$~eV, and a
star-formation rate history of GRBs which is larger than that inferred
from the blue-UV luminosity density.  The cutoff energy for the
galactic-halo component is $E_{max}^{halo}=10^{17.07}$~eV.  The
transition between galactic and extragalactic CRs is found in the
vicinity of the second knee at $\approx 10^{17.6}$~eV, consistent with
a heavy-to-light composition change at this energy.  The ankle, at
$\approx 10^{18.5}$~eV, is interpreted as a suppression from
photo-pair losses \cite{ber04}, analogous to the GZK suppression.

\begin{figure}[t]
\vskip-0.1in
\hskip0.75in\includegraphics[scale=0.4]{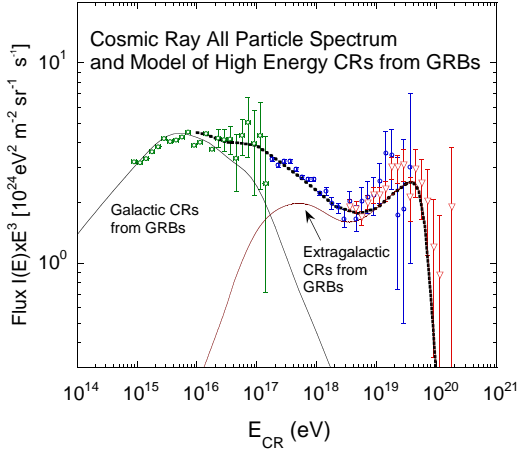}     
\caption{Fit to the all-particle high-energy 
cosmic ray spectrum for a model where GRBs accelerate cosmic rays
\cite{wda04}.  }
\end{figure}


By normalizing the energy injection rate to that required to produce
the CR flux from extragalactic sources observed locally, we determine
the amount of energy a typical GRB must release in the form of
nonthermal hadrons. Our results imply that GRB blast waves are
baryon-loaded by a factor $f_{CR}\gtrsim 60$ compared to the primary
electron energy that is inferred from the fluxes of hard X-rays and
soft $\gamma$ rays measured from GRBs.  For the large baryon load
required for the proposed model of HECRs, calculations show that 100
TeV -- 100 PeV neutrinos could be detected several times per year from
all GRBs with kilometer-scale neutrino detectors such as IceCube
\cite{da03,wda04}.  Detection of even 1 or 2 neutrinos from GRBs with
IceCube or a northern hemisphere neutrino detector will provide
compelling support for this scenario for the origin of high-energy and
UHE cosmic rays.


If GRBs are the sources of HECRs, then high-energy neutrons will be
formed at the burst site through photo-pion processes and, being
neutral, can escape to intergalactic space.  The decay of the neutrons
far from the GRB through the process $n\rightarrow p + e^- +\bar
\nu_e$ leads to $\beta$-decay electrons that make weak synchrotron and
Compton radiation. The best prospect for discovering neutron-decay
halos is to search for diffuse optical synchrotron halos surrounding
field galaxies that display active star formation \cite{der02}. GRBs
that have recently taken place in our Galaxy will produce Compton
emissions at TeV energies that could be detected by the {\it HESS} and
{\it VERITAS} imaging air Cherenkov telescopes
\cite{ikm04}. Nonthermal synchrotron and Compton radiations produced
by secondaries formed in photopion processes by UHECRs traveling
through intergalactic space will also form a nonthermal component of
the diffuse radiation background, which can be used to measure the
magnetic field of intergalactic space.


Because the Milky Way is actively making young high-mass stars, GRBs
will also occur in our Galaxy. The rate of GRBs in the Milky Way is
very uncertain because of lack of precise knowledge about the opening
angle of GRB jets, but could be as frequent as once every 10,000
years. Over the age of the Galaxy, there is a good chance that a
nearby powerful GRB with a jet oriented towards Earth could have
lethal consequences for life. It has recently been argued \cite{mel04}
that such an event contributed to the Ordovician extinction event 440
Myrs ago. UHECRs accelerated by galactic cosmic rays will produce an
additional radiation hazard \cite{dh05}, though we predict that GRBs
cannot produce point-sources of UHECRs.

\section{Conclusions}

GRBs are established to be sources of high-energy $\gamma$ rays.  Here
we have argued GRBs also accelerate hadrons, and that GRBs are sources
of high-energy cosmic rays. Evidence for this is suggested by the slow
decay of $\gamma$ rays from GRB 941017, and by the anomalous
$\gamma$-ray emission component in GRB 941017.

Several types of observations can test this hypothesis. Most
unambiguous is the detection of high-energy neutrinos from a GRB,
which would require an ultra-relativistic hadronic component that is
much more powerful than the nonthermal electron component that
produces the hard X-ray and soft $\gamma$-ray emissions from GRBs
\cite{da03}. Another prediction is the detection of hadronic emission
components in the spectra of GRBs, as observed in GRB 941017
\cite{gon03}. A third observation that would implicate GRBs as the
sources of HECRs is the detection of high-energy neutron $\beta$-decay
halos around star-forming galaxies \cite{der02}, including
$\beta$-decay emission from GRBs in the Milky Way \cite{ikm04}.

\vskip0.1in
I would like to acknowledge joint research with Armen Atoyan, James Chiang, 
Jeremy Holmes, and Stuart Wick.  This work was supported by the Office of
Naval Research and 
NASA GLAST grant DPR S-13756G.

%
\label{Dermer}
 
\end{document}